\documentclass[aps,pra,reprint,superscriptaddress,longbibliography,floatfix]{revtex4-1}

\usepackage{graphicx}% Include figure files.

\graphicspath{ {Figs/} }
\usepackage{dcolumn}% Align table columns on decimal point.
\usepackage{bm}% Bold math.
\usepackage{amsmath}%Allows vphantom command.

\usepackage{lipsum}
\usepackage{appendix}
\usepackage{gensymb}%Allows for the \degree command when writing the degree symbol.

\usepackage{lineno}
\usepackage{hyperref} %hyperref should be last
\hypersetup{breaklinks=true}

 \newcommand{\eh}{\mbox{e-h}\ }%To prevent linebreaks in the middle of this

\begin{document}

\setpagewiselinenumbers
\modulolinenumbers[1]
%\linenumbers
\def\linenumberfont{\normalfont\tiny\sffamily}

\title[Physical Review Applied]{Mapping Charge Recombination and the Effect of Point Defect Insertion in Gallium Arsenide Nanowire Heterojunctions}

%\title[ACS Nano]{Mapping Carrier Recombination at Surface and Point Defects in Gallium Arsenide Nanowire Heterojunctions}
%\title[ACS Nano]{Mapping the Electronic Impact of Surface Defects and Point Defect Insertion in Nanowire Heterojunctions}
%\thanks{Footnote to title of article.}
%%%%%%%%%%%%%%%%%%%%%%%%%%%%%%%%%%%%%%%%%%%%%%%%%%%%%%%%%%%%%%%%%%%%%
%% Some journals require a list of abbreviations or keywords to be
%% supplied. These should be set up here, and will be printed after
%% the title and author information, if needed.
%%%%%%%%%%%%%%%%%%%%%%%%%%%%%%%%%%%%%%%%%%%%%%%%%%%%%%%%%%%%%%%%%%%%%


%%%%%%%%%%%%%%%%%%%%%%%%%%%%%%%%%%%%%%%%%%%%%%%%%%%%%%%%%%%%%%%%%%%%%
%% The manuscript does not need to include \maketitle, which is
%% executed automatically.
%%%%%%%%%%%%%%%%%%%%%%%%%%%%%%%%%%%%%%%%%%%%%%%%%%%%%%%%%%%%%%%%%%%%%

\author{Brian Zutter}
\affiliation{Department of Physics \& Astronomy and California NanoSystems Institute, University of California, Los Angeles, California 90095 USA}
\author{Hyunseok Kim}
\affiliation{Department of Electrical Engineering, University of California, Los Angeles, California 90095 USA }
\author{William Hubbard}
\affiliation{Department of Physics \& Astronomy and California NanoSystems Institute, University of California, Los Angeles, California 90095 USA}
\author{Dingkun Ren}
\affiliation{Department of Electrical Engineering, University of California, Los Angeles, California 90095 USA  }
\author{Matthew Mecklenburg}
\affiliation{Core Center of Excellence in Nano Imaging (CNI), University of Southern California, Los Angeles, California, 90089 USA}
\author{Diana Huffaker}
\affiliation{Department of Electrical Engineering, University of California, Los Angeles, California 90095 USA  }
\affiliation{School of Physics and Astronomy, Cardiff University, Cardiff CF24 3AA, United Kingdom}
\author{B. C. Regan}\email{regan@physics.ucla.edu }
\affiliation{Department of Physics \& Astronomy and California NanoSystems Institute, University of California, Los Angeles, California 90095 USA}

\date{\today}% It is always \today, today,
%  but any date may be explicitly specified

%%%%%%%%%%%%%%%%%%%%%%%%%%%%%%%%%%%%%%%%%%%%%%%%%%%%%%%%%%%%%%%%%%%%%
%% The "tocentry" environment can be used to create an entry for the
%% graphical table of contents. It is given here as some journals
%% require that it is printed as part of the abstract page. It will
%% be automatically moved as appropriate.
%%%%%%%%%%%%%%%%%%%%%%%%%%%%%%%%%%%%%%%%%%%%%%%%%%%%%%%%%%%%%%%%%%%%%
%\begin{tocentry}
%
%Some journals require a graphical entry for the Table of Contents.
%This should be laid out ``print ready'' so that the sizing of the
%text is correct.
%
%Inside the \texttt{tocentry} environment, the font used is Helvetica
%8\,pt, as required by \emph{Journal of the American Chemical
%Society}.
%
%The surrounding frame is 9\,cm by 3.5\,cm, which is the maximum
%permitted for  \emph{Journal of the American Chemical Society}
%graphical table of content entries. The box will not resize if the
%content is too big: instead it will overflow the edge of the box.
%
%This box and the associated title will always be printed on a
%separate page at the end of the document.
%
%\end{tocentry}

%%%%%%%%%%%%%%%%%%%%%%%%%%%%%%%%%%%%%%%%%%%%%%%%%%%%%%%%%%%%%%%%%%%%%
%% The abstract environment will automatically gobble the contents
%% if an abstract is not used by the target journal.
%%%%%%%%%%%%%%%%%%%%%%%%%%%%%%%%%%%%%%%%%%%%%%%%%%%%%%%%%%%%%%%%%%%%%

\begin{abstract}

Electronic devices are extremely sensitive to defects in their constituent semiconductors, but locating electronic point defects in bulk semiconductors has previously been impossible. Here we apply scanning transmission electron microscopy (STEM) electron beam-induced current (EBIC) imaging to map electronic defects in a GaAs nanowire Schottky diode.  Imaging with a non-damaging 80 or 200~kV STEM acceleration potential reveals a minority-carrier diffusion length that decreases near the surface of the hexagonal nanowire, thereby demonstrating that the device's charge collection efficiency (CCE) is limited by surface defects. Imaging with a 300~keV STEM beam introduces vacancy-interstitial (VI, or Frenkel) defects in the GaAs that increase carrier recombination and reduce the CCE of the diode.  We create, locate, and characterize a single insertion event, determining that a defect inserted 7~nm from the Schottky interface broadly reduces the CCE by $10\%$ across the entire nanowire device. Variable-energy STEM EBIC imaging thus allows both benign mapping and pinpoint modification of a device's \eh recombination landscape, enabling controlled experiments that illuminate the impact of both extended (1D and 2D) and point (0D) defects on semiconductor device performance.

\end{abstract}

\maketitle

%%%%%%%%%%%%%%%%%%%%%%%%%%%%%%%%%%%%%%%%%%%%%%%%%%%%%%%%%%%%%%%%%%%%%
%% Start the main part of the manuscript here.
%%%%%%%%%%%%%%%%%%%%%%%%%%%%%%%%%%%%%%%%%%%%%%%%%%%%%%%%%%%%%%%%%%%%%

Crystal defects in semiconductor devices, whether present at fabrication or introduced later via radiation damage, can dramatically impair device performance\cite{2010Johnston,2018McCluskey,2016Hieckmann,srour_displacement_2013,li_radiation_2017, hirst_intrinsic_2016}.
Commonly-used methods for characterizing semiconductor defects have spatial resolution that is crude compared to the feature size in modern microelectronic devices. For example, capacitance-voltage (CV) profiling \cite{thomas_impurity_1962} and deep-level transient spectroscopy (DLTS)\cite{lang_deeplevel_1974} can extract defect concentrations and energy levels, respectively, from simple heterojunctions. But the spatial information provided by these techniques is one dimensional at best. Two dimensional mapping is possible with scanning electron microscope electron-beam induced current (SEM EBIC) imaging, which can locate electrically-active extended (i.e. one- and two-dimensional) defects\cite{collins_proton_2017,warner_displacement_2007,naumann_explanation_2014,2016Hieckmann}, monitor the development of conducting filaments in metal-oxide resistive memory\cite{hoskins_stateful_2017}, measure depletion region widths \cite{ban_direct_2002}, and map minority carrier diffusion lengths\cite{gutsche_direct_2012, chang_electrical_2012,2016Hieckmann,2008Allen}. However, the spatial resolution of SEM EBIC imaging is limited by the size of its \eh (electron-hole) generation  volume\cite{donolato_contrast_1979}. In a standard, electron-opaque SEM sample, primary (beam) electrons deposit most of their energy near the end of their range. The resulting pear-shaped \eh generation volumes  are of order $100$~nm on a side\cite{donolato_contrast_1979,1998Reimer}, which is large compared to feature sizes in many modern devices.

Because a STEM sample is electron-transparent, the corresponding \eh generation volume is the cylindrical, narrow neck of the SEM \eh generation pear\cite{kociak_cathodoluminescence_2017}. With this much smaller \eh generation volume STEM EBIC imaging has the potential to achieve much higher spatial resolution than SEM EBIC imaging\cite{white_imaging_2015, poplawsky_nanoscale_2016, han_interface-induced_2014, warecki_measuring_2018, hubbard_stem_2018,2019Mecklenburg}. Moreover, the higher beam energies accessible with STEM (usually 60--300~keV vs.\ the 1--30~keV of SEM) span the knock-on threshold in semiconductors, which allows a STEM operator to choose whether or not to introduce knock-on displacements in a semiconductor device \emph{precisely} at the position of the sub-nm$^2$ STEM beam. The combination of superior spatial resolution and precision modification enables \emph{in situ} STEM EBIC experiments that directly reveal \eh recombination physics in semiconductor nanodevices. In essence, the STEM's focused electron beam serves both as a highly localized source of $\beta$-radiation damage, and as an immediate local probe of its effects.  This combination allows individual point (i.e. zero-dimensional) defects to be located to within $<1$~nm$^2$.

\begin{figure*}%The asterisk allows you to put the figure the width of the page.
	\includegraphics[width=\linewidth]{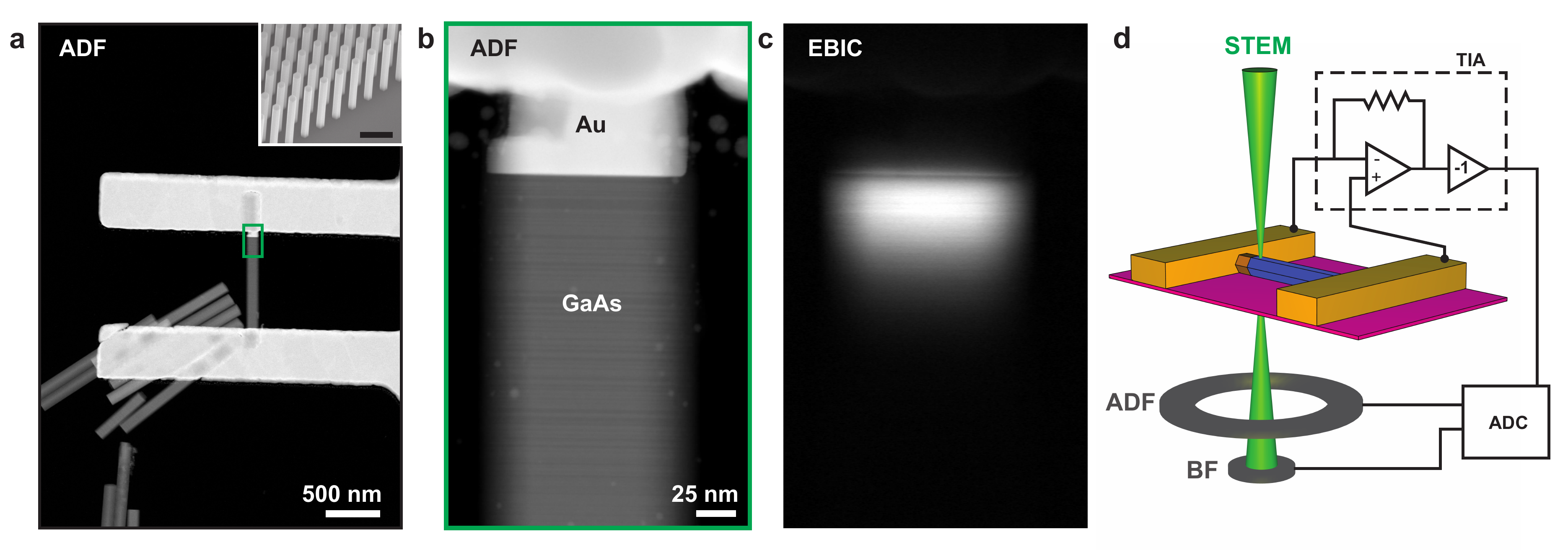}
	\caption{\textbf{STEM EBIC imaging of a Au-GaAs nanowire heterojunction}. A low-magnification, 200~kV STEM annular dark-field (ADF) image (a) of a device shows 130~nm diameter GaAs nanowires and 250~nm-thick, lithographically-defined gold contacts supported by a 15~nm-thick silicon nitride membrane. An SEM image (a, inset) acquired with 30$\degree$ stage tilt, shows the nanowires as grown, before transfer to the silicon nitride membrane (unlabeled scale bar is 500 nm). When the region indicated in green in (a) is imaged at higher-magnification (b), twin boundaries in the GaAs are apparent. An EBIC image (c), acquired simultaneously with (b), reveals the \eh separation that occurs near the Au-GaAs heterojunction. The electrical connections and the locations of the TIA, STEM detectors (ADF,BF), and analog-to-digial converter (ADC) are indicated on a cartoon (d).}
	\label{fig:FigDeviceLayout}
\end{figure*}

To produce targets for demonstrating these capabilities, we fabricate heterojunctions in semiconductor nanowires (Fig.~\ref{fig:FigDeviceLayout}a), which are model systems for elucidating defect physics\cite{li_radiation_2017,gutsche_direct_2012,joyce_phase_2010, dick_crystal_2010,rudolph_spontaneous_2013, chang_electrical_2012,2008Allen}. We put Au contacts on 130~nm-diameter p-type GaAs nanowires (Fig.~S1) with electron-beam lithography, and then briefly anneal the devices \cite{orru_formation_2014,  fauske_situ_2016} (see Supplementary Information). At elevated temperatures gallium and arsenic interdiffuse with the gold at the contacts, forming abrupt ($<2$~nm) axial Au-GaAs heterojunctions aligned with the (111) GaAs planes (Fig.~\ref{fig:FigDeviceLayout}b). Since the growth direction of the GaAs nanowires is along the [111] crystalline direction, these heterojunctions are self-aligned perpendicular to the nanowire axis.  Striations in STEM annular dark-field (ADF) images of the GaAs nanowire (Fig.~\ref{fig:FigDeviceLayout}b) indicate twin boundaries within the zincblende crystal\cite{joyce_phase_2010}.

\begin{figure}%The asterisk allows you to put the figure the width of the page.
	\includegraphics[width=8 cm]{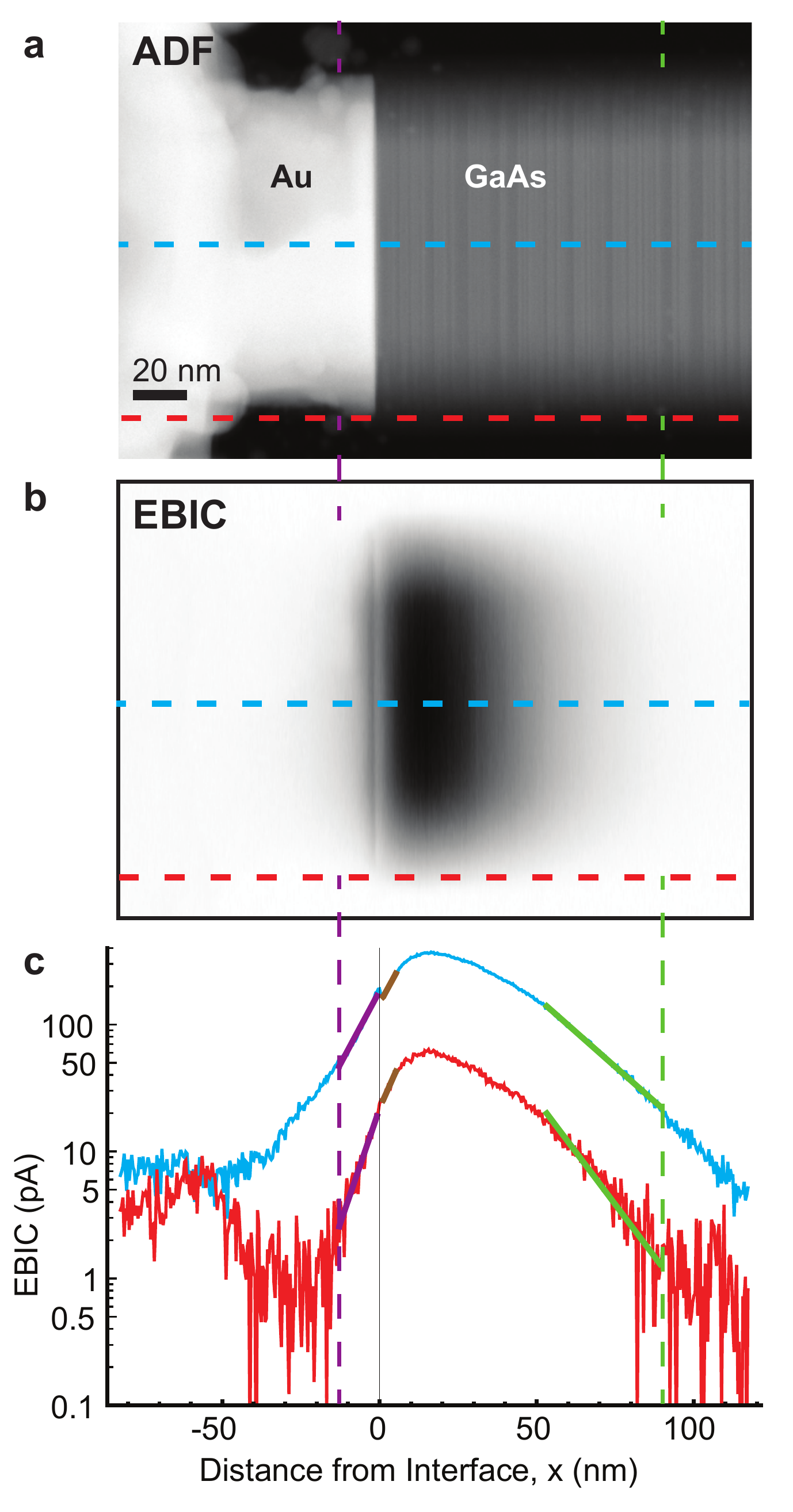}
	\caption{\textbf{Mapping \eh recombination \emph{along} the nanowire.} A STEM ADF image (a)  and  EBIC image (b) are acquired simultaneously with a 200~kV accelerating potential. EBIC line profiles (c) are extracted from the center (dashed blue line) and the edge (dashed red line) of the device in (a,b). The EBIC decay length in the gold (purple line) and the GaAs (brown line) measures the radius $R$ of the \eh generation volume $G$ in each material. In the GaAs far ($> 50$~nm) from the interface, the EBIC decay length (green line) measures the minority-carrier diffusion length $L$. All panels are aligned on the same $x$ axis.}
	\label{fig:FigSingleProfile}
\end{figure} 

Connecting a transimpedance amplifier to a device (Fig.~\ref{fig:FigDeviceLayout}) allows us to generate a STEM EBIC image simultaneously with every STEM ADF image \cite{white_imaging_2015,hubbard_stem_2018,2019Mecklenburg}. The contrast mechanisms generating the two types of images are entirely different, and thus the images provide complementary information. The ADF images (Fig.~\ref{fig:FigDeviceLayout}a,b) provide information only about the device's physical structure (e.g. composition and crystal lattice orientation), while the EBIC image (Fig.~\ref{fig:FigDeviceLayout}c) also reveals the device's electronic structure, in this case the CCE, the size of the space-charge region, and the minority carrier diffusion length.

We understand the EBIC signal as being generated as follows.  Within some generation volume $G$ surrounding the path of the primary electrons through the sample, the STEM electron beam creates \eh pairs.  The pairs are predominantly created by plasmon decay in the GaAs, where plasmons can either be directly created by primary electrons in the GaAs or by secondary electrons resulting from primary electrons in the GaAs or the Au.\cite{1998Reimer,kociak_cathodoluminescence_2017}. Electrons in the GaAs conduction band, which are the minority carriers in these p-doped nanowires, then diffuse some distance, parametrized by the minority carrier diffusion length, before recombining probabilistically. Electrons that happen to diffuse to the space-charge region near the Au-GaAs heterojunction can be permanently separated from their holes by the built-in electric field $E$. The separated charge is collected by the electrodes and constitutes the EBIC. The CCE,  which here is the ratio of the EBIC to the rate of \eh pair generation, determines what fraction of \eh pairs are collected. The \eh generation rate is relatively insensitive to crystal defects, while the CCE is lowered by recombination centers within the GaAs. Thus relative changes in CCE due to carrier recombination have a proportional effect on the EBIC. The EBIC also depends on the beam position within the nanowire through three size scales: the radius $R$ of the \eh generation volume, the minority-carrier diffusion length $L$, and the thickness $t$ of the space-charge region.  A single EBIC image can provide information on each of them\cite{nichterwitz_numerical_2013}.

Imaging another device with STEM ADF (Fig.~\ref{fig:FigSingleProfile}a) and STEM EBIC (Fig.~\ref{fig:FigSingleProfile}b) shows how these length scales collectively determine the shape of the EBIC profile (Fig.~\ref{fig:FigSingleProfile}c). The STEM ADF image shows the location of the heterojunction, twin boundaries in the GaAs, and some voiding in the Au. The STEM EBIC image shows a CCE that varies in a non-trivial way as a function of position. Just as an optical point-spread function limits the resolution of an optical microscope, the size of the \eh generation volume, $G$, limits the EBIC electronic spatial resolution. It manifests itself clearly here in at least two ways.  First, a non-zero EBIC is generated when the beam is incident on the Au side of the heterojunction, even though Au is not a semiconductor and has no band gap. Fitting an EBIC line profile from the center of the nanowire (Fig. \ref{fig:FigSingleProfile}c, blue profile, purple line) to an exponential $I \propto e^{x/R}$ yields a decay length $R = 9.4 \pm 0.2$~nm, where the error bar reflects the statistical uncertainty in a linear least-squares fit. This length scale measures how far secondary electrons can travel in the gold and still create \eh pairs in the GaAs, i.e. the radius $R$ of $G$ within the Au. Second, the EBIC profile maximum 20~nm away from the heterojunction interface indicates that, near the heterojunction, $G$ in the GaAs is truncated by the Au. Fitting the EBIC along the center-line in the GaAs immediately adjacent to the heterojunction (blue profile, brown line) to an exponential $I \propto e^{x/R}$ yields $R = 9.6 \pm 0.4$~nm, which indicates the radius $R$ of $G$ within the GaAs. This model also explains the hiccup in the line profiles (also clearly visible in Fig.~\ref{fig:FigSingleProfile}b) at the heterojunction: moving across the boundary into the Au actually increases the EBIC (even though the Au does not support \eh pairs) because $G$ is continuous while the absolute number of secondary electrons increases discontinuously.

On the Au side of the heterojunction the electric field $E=0$, while on the GaAs side a substantial electric field $E \ne 0$ exists in the space-charge region. Thus near the heterojunction the CCE is a step function with approximate values of zero within the Au and unity within the GaAs \cite{nichterwitz_numerical_2013}, and the EBIC $\propto G \times $CCE measures $G$ as just described.  Far ($> 50$~nm) from the heterojunction in the GaAs the $E$-field returns to zero, the minority-carrier transport is dominated by diffusion, and the EBIC measures the CCE.  With increasing distance from the space charge region the EBIC in the GaAs decays exponentially, with a decay length equal to the minority-carrier diffusion length $L$. Fitting the EBIC current in the center of the nanowire (Fig. \ref{fig:FigSingleProfile}c, blue profile) to $I \propto e^{-x/L}$, where $x$ is the distance from the heterojunction, gives $L =19.7 \pm 0.1$~nm (green line), where the error bar again reflects the statistical uncertainty in a linear least-squares fit. This relatively short diffusion length likely results not from the nanowire's dense zincblende twin boundaries (Fig~\ref{fig:FigSingleProfile}a), but rather from surface recombination \cite{chang_electrical_2012}.  For instance, the surface-to-volume ratio at the thin edge of the nanowire is larger, and an EBIC profile at the edge (Fig.~\ref{fig:FigSingleProfile}c, red profile) shows a much smaller minority-carrier diffusion length  $L = 13.2 \pm 0.9$~nm (green line). Thus \eh pairs generated nearer the nanowire surface are more likely to recombine.  While $L$ is much smaller than the nanowire diameter $D=130$~nm, this fact is not as surprising as it might seem at first: on average, any point in a long cylinder of diameter $D$ is only a distance $D/6$ away from the cylinder surface.

\begin{figure}%The asterisk allows you to put the figure the width of the page.
	\includegraphics[width=9 cm]{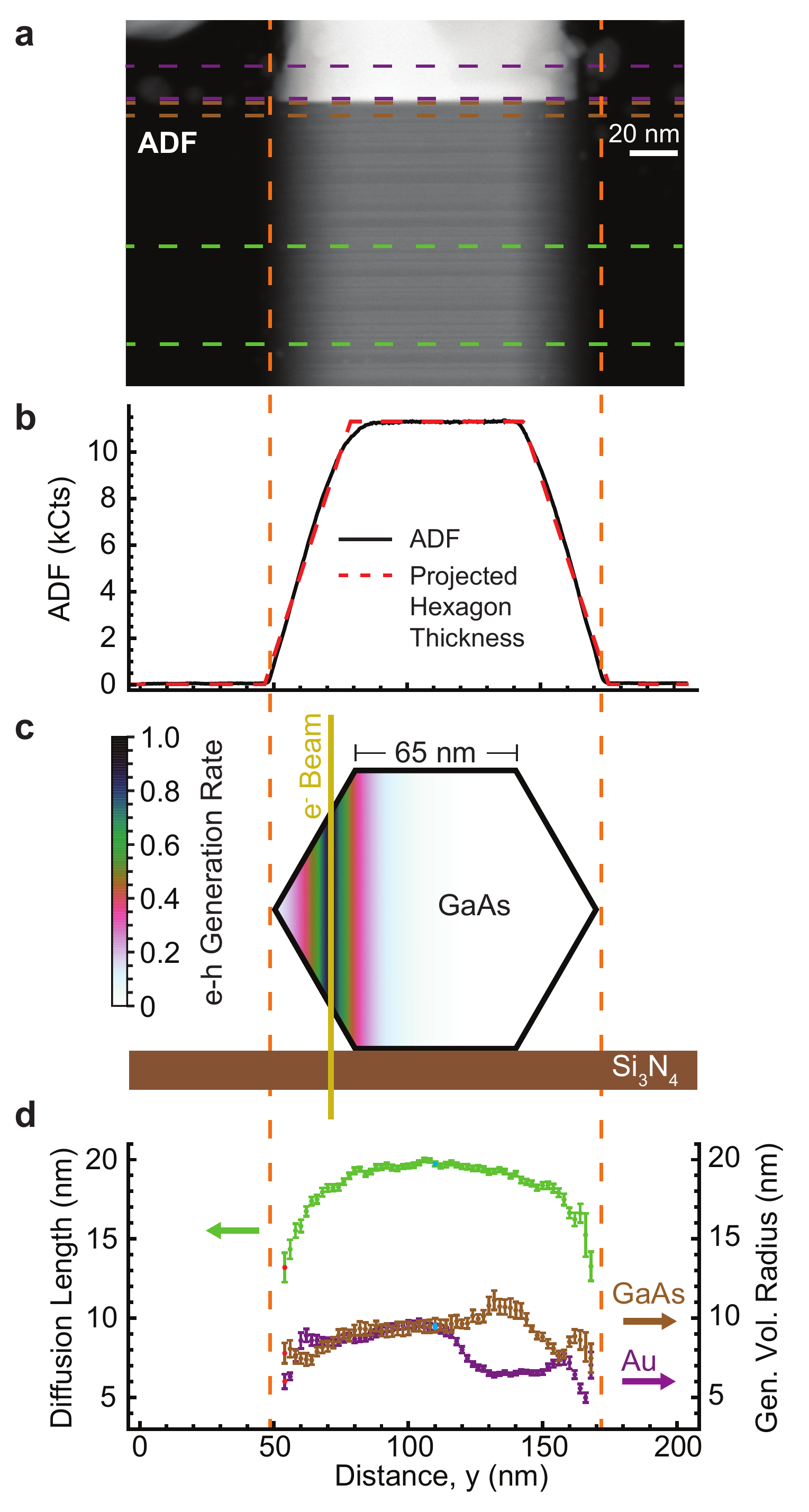}
	\caption{\textbf{Mapping \eh recombination \emph{across} the nanowire.} (a) The data of Fig. \ref{fig:FigSingleProfile}a are rotated $90\degree$  to align the Schottky interface with the horizontal axis. Summing the ADF signal from the dashed-green ROI in (a) gives a profile (b) approximately proportional to the sample thickness. This profile agrees well with the projected thickness of a geometrically-perfect hexagon (dashed-red line in b). A slice of the cylindrical \eh generation volume is overlaid on the nanowire cross-section (c), with a decay radius $R = 10$~nm. The generation volume radii $R$ in gold and GaAs and the minority-carrier diffusion length $L$ in the GaAs are plotted as a function of radial position across the hexagonal nanowire in (d). The $L$ and $R$  measurements shown explicitly in Fig.~\ref{fig:FigSingleProfile}d are highlighted in blue (center) and red (edge) here.   All panels are aligned horizontally on the same distance axis.} %The minority-carrier diffusion length decreases dramatically near the nanowire's narrow edges due to increased surface recombination.
	\label{fig:FigDiffusionMap}
\end{figure}

\begin{figure*}%The asterisk allows you to put the figure the width of the page.
	\includegraphics[width= \linewidth]{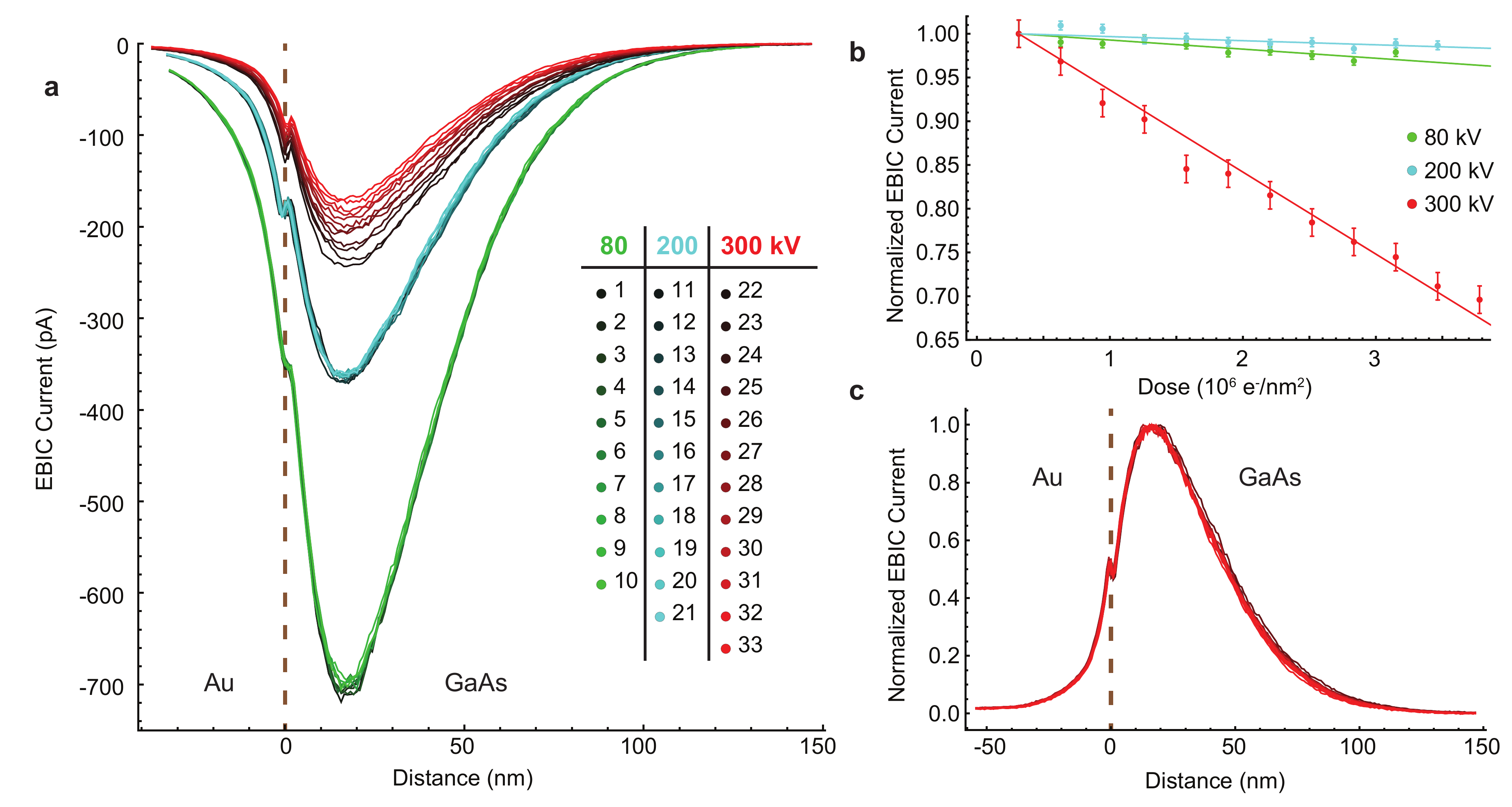}
	\caption{\textbf{STEM EBIC at 80, 200, and 300~kV accelerating voltage}. Line profiles (a) show the effect of repeated imaging of a device (shown in Fig. \ref{fig:DFRadDamage}) at 80~kV (green), 200~kV (cyan) and 300~kV (red). Line profiles are extracted from the cyan boxes shown in Fig. \ref{fig:DFRadDamage}a. Only the 300~kV curves show a significant decrease in the EBIC with repeated imaging. Plotting the profile minima, normalized relative to their initial values, versus dose shows (b) a linear dose effect at 300~kV and insignificant effects at 80 and 200~kV. At 300~kV the maximum EBIC decreases by $\sim 3$\% per image. Error bars on the 80, 200 and 300~kV data series are determined by setting the reduced $\chi^2=1$ for the linear fits. Normalizing the profiles of the 300~kV data series by the minimum value of each (c) shows that only the amplitude of the EBIC line profile changes, not the shape.}
	\label{fig:FigProfileskV}
\end{figure*}

The nanowire's simple shape facilitates the interpretation of the EBIC data. ADF STEM data (Fig.~\ref{fig:FigDiffusionMap}a $=$ Fig.~\ref{fig:FigSingleProfile}a rotated) show that the nanowire's cross-section (Fig.~\ref{fig:FigDiffusionMap}b) is a near-perfect hexagon (Fig.~\ref{fig:FigDiffusionMap}c). To give a sense of scale, a slice of a cylindrical \eh generation volume with $R = 10$~nm is superimposed on the GaAs nanowire's hexagonal cross section in Fig. \ref{fig:FigDiffusionMap}c. 

STEM EBIC imaging's extraordinary spatial resolution reveals how charge recombination varies as a function of not only the nanowire's axial coordinate, but also its radial coordinate (compare e.g. Ref.~\onlinecite{chang_electrical_2012}). Extending the STEM EBIC analysis of Fig.~\ref{fig:FigSingleProfile} by fitting at every distinct axial coordinate, we  map both the minority-carrier diffusion lengths $L$ (which determine the CCE) and the radii $R$ (which determine $G$) across the width of the nanowire. The diffusion length $L$ decreases from 20~nm  near the center axis of the 130~nm-wide nanowire to $13$~nm near the  edges (Fig.~\ref{fig:FigDiffusionMap}d, green plot), as expected for recombination occurring primarily at the nanowire surface.

The generation volume $G$'s effective radius $R$ is less than 10~nm in both  the gold (Fig. \ref{fig:FigDiffusionMap}d, purple points) and the GaAs (brown points). Due to voiding in the Au (see Figs.~\ref{fig:FigSingleProfile}a and \ref{fig:FigDiffusionMap}a), these curves are irregular on one side of the nanowire.     Near the nanowire's center the corresponding (cylindrical) STEM EBIC generation volume is $G \sim 4 \times 10^4$~nm$^3$, while an SEM generation volume with effective radius $r \simeq 100$~nm (appropriate for a 5~keV accelerating voltage\cite{donolato_contrast_1979}) is $\times 100$ larger. STEM EBIC's resolution advantage is $\sim r/R$, or a factor of 10, relative to SEM EBIC.  

Taking the `electronic structure resolution' to be the full-width, half-maximum (FWHM) of the generation volume, our measured resolution is $(2 \ln 2) R=14$~nm.  Note that our STEM EBIC images show smaller features, implying better STEM EBIC resolution, but that these features are primarily generated by changes in physical structure, not electronic structure.  For instance, the STEM EBIC images show both the thickness variations that accompany the twin boundaries ($\sim 2$~nm) and the EBIC hiccup ($\sim 3$~nm) at the heterojunction (Fig.~S6).

This resolution advantage creates qualitatively new capabilities: STEM EBIC, unlike SEM EBIC, can map device parameters like the minority-carrier diffusion length \emph{across} an individual nanowire. Our measured $R$ of 10~nm is an order-of-magnitude larger than predicted by the CASINO Monte Carlo simulator\cite{kociak_cathodoluminescence_2017, hovington_casino_1997, drouin_casino_2007}. We attribute this discrepancy to CASINO's omission of plasmon generation (the dominant energy loss mechanism in GaAs for electrons of $<50$~eV energy \cite{rothwarf_plasmon_1973}) in its calculation of stopping power at low electron energies.

Within 50~nm of the interface, the EBIC signal is below the continuation of the green lines on the log-linear plot (Fig.~\ref{fig:FigSingleProfile}c).  Near the heterojunction we might instead expect the EBIC increase as the local $E$-field increases in the space-charge region. As mentioned above, the observed decrease indicates that some of the \eh generation volume $G$ is in the Au (Fig~\ref{fig:FigDiffusionMap}d). Thus the EBIC data indicates that the thickness $t$ of the space-charge region is less than the radius $R < 10$~nm of the \eh pair generation volume $G$.

To compare damage rates at various accelerating voltages, we image the device of Figs. \ref{fig:FigSingleProfile} and \ref{fig:FigDiffusionMap} while keeping all other imaging conditions (e.g. the 50~pA STEM beam current, 762~$\mu$s pixel dwell time, and 0.87~nm pixel size) constant (Fig.~\ref{fig:FigProfileskV}). Repeated imaging at 80~kV and 200~kV has little effect on the EBIC, but 300~kV imaging markedly reduces the EBIC signal (Fig.~\ref{fig:FigProfileskV}a). (The EBIC magnitude decreases as the accelerating potential increases because higher energy electrons deposit less energy per distance traveled in a solid \cite{egerton_EELSbook_2011}.)

\begin{figure*}
	\includegraphics[width=\linewidth]{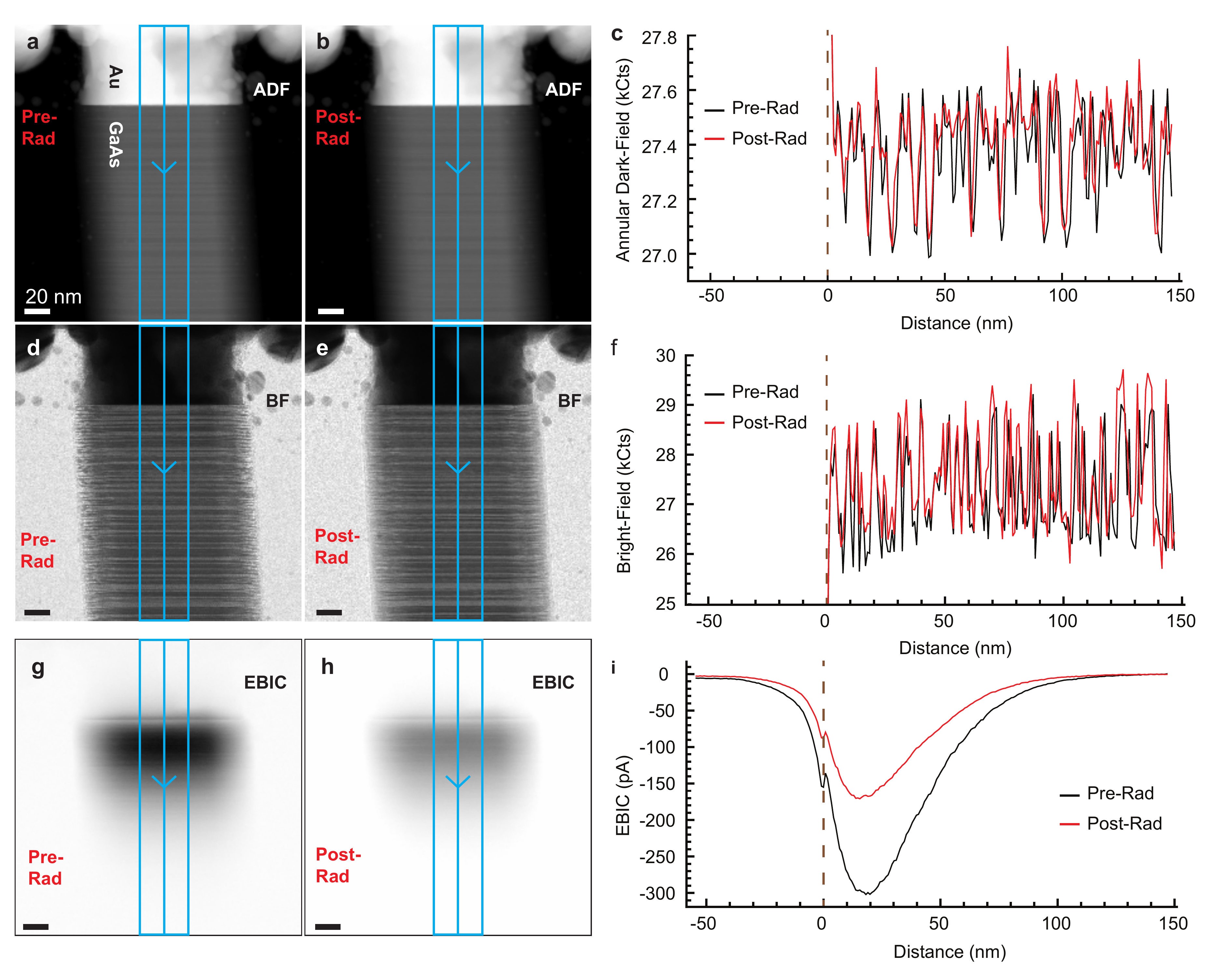}
	\centering
	\caption[BeforeAfterDF]{  \textbf{Annular dark-field, bright-field, and EBIC imaging before and after irradiation with 300~kV STEM electrons.} STEM ADF, bright-field (BF), and STEM EBIC images acquired before (a,d,g) and after (b,e,h) a dose of $6.0 \times 10^6$ $e^-/$nm$^2$ at 300~kV accelerating voltage. The total dose is applied while acquiring the twelve images $\#22$--$\#33$ (Fig. \ref{fig:FigProfileskV}) and three alignment images (between $\#21$ and $\#22$). Line profiles are extracted (c,f,i) by horizontally averaging data within the blue boxes.  A dashed brown line in the line profiles indicates the Au-GaAs interface. Irradiation produces almost no change in the conventional imaging channels (ADF, BF), but a $44\%$ decrease in the maximum EBIC, which highlights the advantage of EBIC over conventional imaging for revealing functional properties such as the CCE. }%changes in functional
	\label{fig:DFRadDamage}
\end{figure*}

As a function of dose, the EBIC, and thus the CCE, decreases linearly at 300~kV (Fig.~\ref{fig:FigProfileskV}b). We attribute the reduction in CCE to knock-on damage that introduces electronically-active vacancy-interstitial (VI) defects, probably on the As sublattice \cite{pillukat_point_1996}.  These defects function as \eh recombination centers, reducing the current that is collected to form the EBIC signal.   Energy and momentum conservation dictate that the maximum possible energy transfer from a beam electron to a gallium (mass number $A=70$) nucleus is 2.7, 7.5, and 12.2~eV for incident electron kinetic energies of 80, 200, and 300 keV, respectively \cite{egerton_EELSbook_2011}. The maximum energy transfer varies inversely with the mass of the target nucleus, so the numbers for arsenic (A=75) are nearly the same (2.5,	7.0,	and 11.4~eV, respectively). Gold (A=197) allows only $70/197\sim 1/3$ the energy transfer, which is small enough at all of the accelerating voltages used in these experiments that the displacement or knock-on damage in this material is negligible.  But the displacement damage threshold energy in GaAs is $\sim 10$~eV \cite{2010Johnston,pillukat_point_1996,nordlund_defect_2001, chen_computational_2017} (although with substantial uncertainty --- see Ref.~\cite{chen_computational_2017} and references within), which leads us to expect an onset of electron beam-induced displacement damage between the accelerating voltages of 200 and 300~kV. 

After they have been normalized relative to their minima, all twelve EBIC profiles acquired at the damaging 300~kV accelerating voltage overlap closely (Fig.~\ref{fig:FigProfileskV}c). That the defects introduced do not change the minority-carrier diffusion length $L$  indicates that $L$ is still dominated by surface recombination, and that this length scale is determined by the nanowire cross section as discussed earlier.

\begin{figure*}%The asterisk allows you to put the figure the width of the page.
	\includegraphics[width=\linewidth]{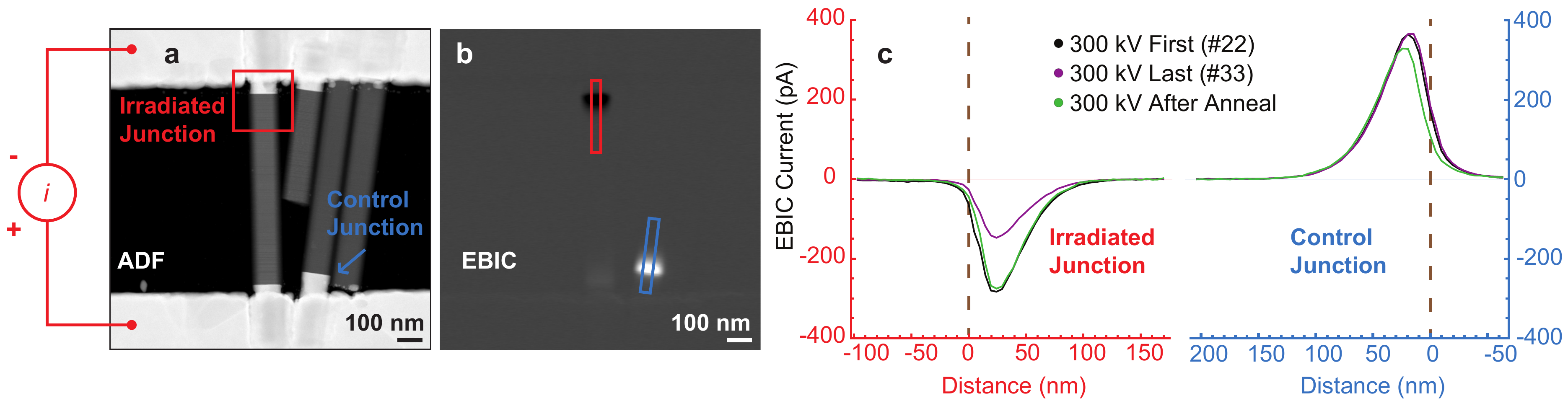}
	\caption{\textbf{STEM EBIC before and after annealing}. A low-magnification STEM ADF image (a) of the device of Figs.~\ref{fig:FigSingleProfile}--\ref{fig:DFRadDamage} shows both the heavily irradiated Au-GaAs heterojunction (red square) and the adjacent control, a heterojunction irradiated less frequently and only at low magnification (cyan arrow). The simultaneously-acquired EBIC image (b) shows that the two heterojunctions have EBICs with opposite signs because of their relative orientations in the circuit. The red and cyan rectangles in (b) indicate the sources of the line profiles in the red and cyan plots of (c).  After irradiation with 300~kV electrons the magnitude of the irradiated heterojunction's EBIC is reduced relative to the control. After an anneal the irradiated heterojunction's EBIC recovers. }
	\label{fig:FigHealing}
\end{figure*}

Repeated imaging of this device at 300~kV  thus causes a substantial reduction in the EBIC (and thus the CCE) of the nanowire junction --- the radiation damage destroys this device's ability to effectively separate of \eh pairs. Given the large dose (six million 300~keV electrons per square nanometer) and accompanying efficiency drop, it is remarkable that the device appears undamaged in the standard STEM imaging channels (Figs.~\ref{fig:DFRadDamage}a--f).  But while standard STEM imaging is blind to the inserted defects, which have a relatively minor effect on the nanowire's physical structure, EBIC imaging (Figs.~\ref{fig:DFRadDamage}g--i), vividly reveals their outsize impact on the nanowire's electronic structure (namely a $44\%$ reduction of the maximum EBIC). % ADF and BF STEM imaging show  the crystal structure of the nanowire apparently unchanged.

The device of Figs.~\ref{fig:FigSingleProfile}--\ref{fig:DFRadDamage} is part of a larger circuit (Fig.~\ref{fig:FigHealing}). At low magnification a second heterojunction, on an adjacent nanowire but also in the circuit, is visible.  The second heterojunction is imaged at lower magnification and less frequently (4.8~nm pixel size, 0.762~$\mu$s dwell time), and is thus subjected to less than 1\% of the radiation dose of the irradiated junction. This adjacent junction can control for changes that are independent of radiation dose.

To corroborate the role of radiation-induced defects in the observed EBIC reduction, after image \#33 of Figs.~\ref{fig:FigProfileskV}--\ref{fig:DFRadDamage} we anneal the nanowire device in an inert argon atmosphere at 250\degree C for 30~minutes.  Such treatment reduces the density of VI defects within the nanowire, since the elevated temperature makes the beam-induced defects mobile, allowing interstitials and vacancies to meet and annihilate\cite{aukerman_annealing_1962,pillukat_point_1996}. After the annealing treatment, we image the nanowire heterojunction again (Fig.~\ref{fig:FigHealing}). The anneal restores the  EBIC to its pre-irradiated value (i.e. restores the 44\% lost) while changing the measured EBIC in the control junction by only a small amount ($< 10 \%$). The post-anneal restoration is consistent with the hypothesis that the radiation-induced CCE reduction is caused by  defects --- specifically VI defects --- that anneal away at high temperature.

\begin{figure*}%The asterisk allows you to put the figure the width of the page.
	\includegraphics[width=\linewidth]{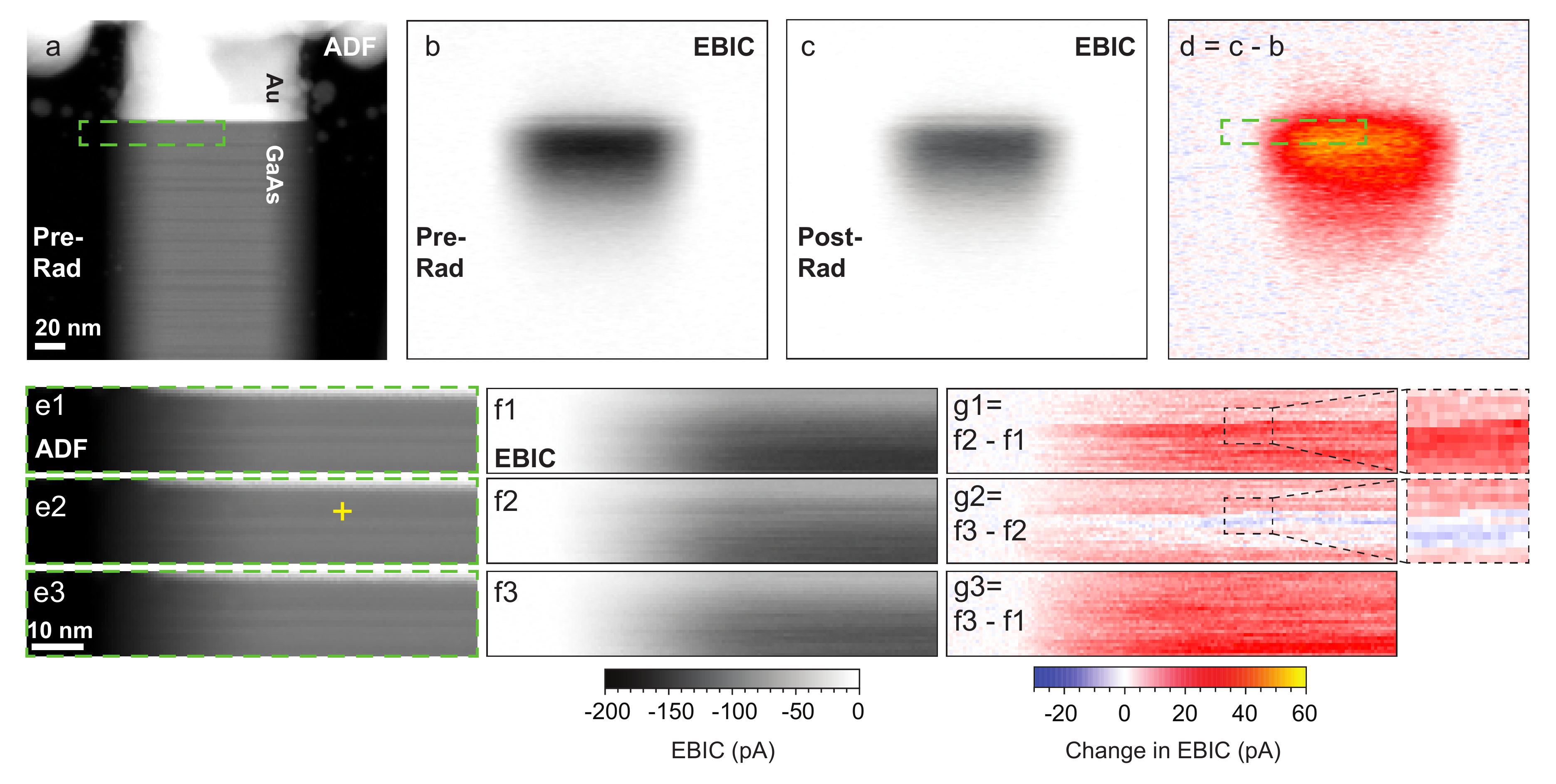}
	\caption{\textbf{Defect insertion and pinpoint localization with STEM EBIC at 300~kV}.  We record the initial state of an Au-GaAs nanowire heterojunction with low-dose ($3.0 \times 10^4$ e$^-$/nm$^2$) ADF STEM (a) and  STEM EBIC (b)  images acquired simultaneously.  We then image the region outlined by the dashed box (a) three times (e1,e2,e3) with a high dose ($1.8 \times 10^6$ e$^-$/nm$^2$ per image). After the three strip images we acquire a second low-dose EBIC image (c). A difference image (d) shows that the EBIC decreases across the entire nanowire, even though the dose was confined to a narrow region on the left side of the nanowire. Dark-field strip images (e1,e2,e3) show no change during irradiation, while the simultaneously-acquired EBIC strip images (f1,f2,f3) show significantly smaller signals. EBIC difference images (g1, g2) reveal a sudden drop in the EBIC magnitude within one 0.63~nm pixel, indicating that a defect was inserted during the second strip image at the location indicated by the yellow cross (e2). Zoom regions (dashed boxes on g1, g2) of 11 pixels $\times 16$~pixels ($7$~nm $\times 10$~nm) demonstrate that both the row and the column of the insertion event can be located precisely. A difference image between the first and third strip image (g3) indicates that, as in (d), the electronic impact of the defect is delocalized. The black-white color scale applies to panels (b,c,f), and the blue-yellow color scale applies to panels d and g.}%as shown in (b) ---
	\label{fig:DefectInsertion}
\end{figure*}

The STEM's precise electron beam positioning allows us to observe the effect of selectively dosing just part of the nanowire.  In an experiment performed on the Fig.~\ref{fig:FigHealing} device (after the annealing experiment), we irradiate a narrow strip of GaAs that only spans half of  the nanowire heterojunction (denoted by dashed green box in Fig.~\ref{fig:DefectInsertion}a).  With 300~kV, a 50~pA beam current, a 0.633~nm pixel size, and a 2.3~ms pixel dwell time, the dose per area per strip image, $1.8 \times 10^6$ e$^-$/nm$^2$, is $5.5\times $ that of the Fig.~\ref{fig:FigProfileskV} experiment. As in the experiment of Fig.~\ref{fig:FigHealing}, we acquire low dose images before and after the high-dose images for purposes of comparison.  (Here a 153~$\mu$s dwell time and 1.27~nm pixel size of the two low-dose images contributes only $1.1\%$ of the combined dose from the three strip images.) The difference between the before and after images (Fig.~\ref{fig:DefectInsertion}d) shows that the localized strip irradiation decreases the CCE across the entire width of the nanowire.

By comparing consecutive EBIC images we can, in some cases, precisely identify the position where an electrically-active defect is inserted. ADF (Fig.~\ref{fig:DefectInsertion} e1, e2, e3) and EBIC (f1, f2, d3) images are collected simultaneously in the three high-dose strip images. In the (standard) raster pattern used here, the electron beam scans across one row from left to right, and then moves down to scan the next rows in sequence in the same direction. Each strip image shows a dose-induced EBIC decrease, as in Fig.~\ref{fig:FigProfileskV}. EBIC difference images (Fig.~\ref{fig:DefectInsertion} g1, g2) reveal a sudden drop (8 pA magnitude) in the EBIC that occurs in a single 0.63~nm pixel. We attribute this sudden drop to the insertion of an electrically-active defect during the second strip image, at the pixel indicated by the yellow cross (Fig.~\ref{fig:DefectInsertion} e2). Notably, since the displaced atom of a VI defect can travel only a few angstroms from its original position at these low energies, and likely in the direction of the electron beam, the yellow cross  marks the final location of this single defect \cite{2010Johnston,chen_computational_2017}. Thus the defect generation volume is much smaller than the \eh generation volume, and EBIC imaging is able to locate VI insertion events with a much higher precision ($<1$~nm) than its electronic resolution of $\approx 10$~nm.

This defect reduces the EBIC magnitude by 11~pA (a $10\%$ reduction), as determined by comparing the mean EBIC of the 10 pixels before the insertion to the mean EBIC of the 10 pixels after the insertion. As with all of the other STEM-beam induced radiation damage here, this insertion leaves no signature in the conventional ADF imaging. The CCE reduction from this individual defect insertion event is again non-local (as in Fig.~\ref{fig:DefectInsertion}d), since the difference between the first and third strip images (Fig.~\ref{fig:DefectInsertion} g3) is uniform.

In summary, STEM EBIC imaging with an electron-beam acceleration potential of 80 or 200~kV maps the CCE of a GaAs nanowire diode without damaging the device. The minority-carrier diffusion length is found to decrease significantly near the thin edges of the nanowire, and is thus limited by surface recombination.  Imaging with the acceleration potential increased to 300~kV introduces defects in the nanowire that decrease the diode's CCE. These VI defects can be annealed away to restore the original CCE of the diode. Despite being invisible in conventional STEM imaging channels, a VI defect inserted at 300~kV can be precisely located by identifying an abrupt drop in CCE as the electron beam rasters.  As these results show, a modern, variable-energy STEM equipped for EBIC imaging is an experimentally potent combination for producing, locating, and characterizing defects in semiconductor devices with high spatial resolution.

\textbf{Methods}: GaAs nanowires are grown by selective-area epitaxy in a vertical metalorganic chemical vapor deposition (MOCVD) reactor (Emcore D-75) at 60~Torr, using hydrogen as a carrier gas. Triethylgallium (TEGa), tertiarybutylarsine (TBAs), and diethylzinc (DEZn) are used as precursors for gallium, arsenic, and zinc p-type dopant, respectively. See supplementary information for complete growth parameters. The GaAs nanowires' measured  resistivity is $\lesssim 5$~$\Omega\cdot$cm (see Fig.~S5 and related text), which in bulk GaAs corresponds to a dopant concentration\cite{sze_resistivity_1968} of $\gtrsim 5 \times 10^{15}$~cm$^{-3}$.

Nanowires are mechanically transferred using a sharp tungsten probe to 15~nm-thick silicon nitride windows reinforced with a 0.8~$\mu$m-thick backing layer of silicon oxide (Fig.~S2). Nanowires are located with a scanning electron microscope (SEM), and individual electron-beam lithography patterns are written to each silicon nitride window using polymethylmethacrylate (PMMA) resist. The samples are dipped in 1:10 hydrofluoric acid:water solution for 60~seconds to remove native GaAs oxides. Immediately afterward, samples are placed in an electron-beam evaporation chamber and 250~nm of gold is deposited. Intruded gold contacts are formed by heating the samples in a rapid thermal annealer (RTA) at 340\degree C for 30~seconds in a nitrogen atmosphere. To make the sample electron-transparent, the silicon oxide support film is removed with a hydrofluoric acid vapor etch. The sample is loaded into a Hummingbird Scientific biasing holder with electrical feedthroughs. STEM images are acquired at 80, 200, and 300~kV accelerating voltage within an FEI Titan STEM. EBIC signal is measured using a FEMTO DLPCA-200 transimpedance amplifier, set to $10^9\,\Omega$ gain with 40~kHz bandwidth. The amplified current signal is fed into an analog input in the STEM, and is synced to the STEM probe position to form an EBIC image.

%%%%%%%%%%%%%%%%%%%%%%%%%%%%%%%%%%%%%%%%%%%%%%%%%%%%%%%%%%%%%%%%%%%%%
%% The "Acknowledgement" section can be given in all manuscript
%% classes.  This should be given within the "acknowledgement"
%% environment, which will make the correct section or running title.
%%%%%%%%%%%%%%%%%%%%%%%%%%%%%%%%%%%%%%%%%%%%%%%%%%%%%%%%%%%%%%%%%%%%%

\textbf{Acknowledgments:} This work was supported by National Science Foundation (NSF) award DMR-1611036, NSF STC award DMR-1548924 (STROBE), and S\^er Cymru grants in Advanced Engineering. Data were collected at the Electron Imaging Center for Nanomachines (EICN) at the California NanoSystems Institute (CNSI). Samples were fabricated in the Integrated Systems Nanofabrication Cleanroom (ISNC) at the CNSI. We thank Ting-Yuan Chang for assistance with patterning the nanowire growth template. %Matthew Mecklenburg for valuable discussions, and

%%%%%%%%%%%%%%%%%%%%%%%%%%%%%%%%%%%%%%%%%%%%%%%%%%%%%%%%%%%%%%%%%%%%%
%% The same is true for Supporting Information, which should use the
%% suppinfo environment.
%%%%%%%%%%%%%%%%%%%%%%%%%%%%%%%%%%%%%%%%%%%%%%%%%%%%%%%%%%%%%%%%%%%%%
%\begin{suppinfo}
%
%A listing of the contents of each file supplied as Supporting Information
%should be included. For instructions on what should be included in the
%Supporting Information as well as how to prepare this material for
%publications, refer to the journal's Instructions for Authors.
%
%The following files are available free of charge.
%\begin{itemize}
%  \item Filename: brief description
%  \item Filename: brief description
%\end{itemize}
%
%\end{suppinfo}

%%%%%%%%%%%%%%%%%%%%%%%%%%%%%%%%%%%%%%%%%%%%%%%%%%%%%%%%%%%%%%%%%%%%%
%% The appropriate \bibliography command should be placed here.
%% Notice that the class file automatically sets \bibliographystyle
%% and also names the section correctly.
%%%%%%%%%%%%%%%%%%%%%%%%%%%%%%%%%%%%%%%%%%%%%%%%%%%%%%%%%%%%%%%%%%%%%
%\textbf{References:} %I'm adding this line because the style file seems to not name the section. bcr
\bibliography{GaAs_NWs}

\end{document}